\documentclass[10pt]{article}
\usepackage{amsmath}
\usepackage{amssymb}
\usepackage{amsfonts}

\begin{document}

\title{Dressed coordinates: the path-integrals approach}
\author{R. Casana\thanks{%
casana@ift.unesp.br}~, G. Flores-Hidalgo\thanks{
gflores@ift.unesp.br}~ and B. M. Pimentel\thanks{
pimentel@ift.unesp.br} \\
%EndAName
\textit{{\small Instituto de F\'{\i}sica Te\'orica, Universidade Estadual
Paulista}} \\
\textit{\small Rua Pamplona 145, CEP 01405-900, S\~ao Paulo, SP, Brazil}}
\date{}
\maketitle

\begin{abstract}
The recent introduced \textit{dressed coordinates} are studied in
the path-integral approach. These coordinates are defined in the
context of a harmonic oscillator linearly coupled to massless scalar
field and, it is shown that in this model the dressed coordinates
appear as a coordinate transformation preserving the path-integral
functional measure. The analysis also generalizes the \textit{sum
rules} established in a previous work.
\end{abstract}

%%%%%%%%%%%%%%%%%%%%%%%%%%%%%%%%%%%%%%%%%%%%%%%%%%%%%%%%%%%%%%%%%%%%%%%%

\section{Introduction}

%%%%%%%%%%%%%%%%%%%%%%%%%%%%%%%%%%%%%%%%%%%%%%%%%%%%%%%%%%%%%%%%%%%%%%%%
In recent works, the concept of dressed coordinates and dressed states have
been introduced in the context of a harmonic oscillator linearly coupled to
a massless scalar field \cite{adolfo1,adolfo2,gabriel,sumrules}. As
emphasized in these references, the introduction of the dressed (or
renormalized \cite{sumrules,yony}) coordinate and state concepts is
necessary in order to give physical consistence to the oscillator-field
system as a toy model to describe an atom in interaction with the
electromagnetic field, where the atom is roughly modeled by the harmonic
oscillator. Also, as early stressed the introduction of dressed coordinates
and states is twofold advantageous. From the physical view point, the
dressed states behave as the one expected for the physically measurable
states: excited atomic states are unstable whereas the atom in their ground
state and no field quanta is stable. On the other hand, it allows exact
computations for the probability amplitudes of the different radiation
processes of the atom. Indeed, when the calculation is performed for weak
coupling constant we obtain, for the spontaneous decay of the first excited
state of the atom, the long know result: $e^{-\Gamma t}$\cite{expodecay}.
Furthermore, when applied to a confined atom, approximated by the harmonic
oscillator, in a spherical cavity of sufficiently small diameter the method
accounts for the experimentally observed inhibition of the decaying
processes \cite{hulet,haroche2}. Besides that, in Refs. \cite{nonlinear,yony}
the extension of the dressed coordinate and state concepts for non linear
system have been addressed.
%%%%%%%%%%%%%%%%%%%%%%%%%%%%%%%%%%%%%%%%%%%%%%%%%%%%%%%%%%%%%%%%%%%%%%%%%

Nevertheless, in all previous works the approach used has been via the
operatorial formalism of Quantum Mechanics. The aim of this paper is to
develop a path-integral approach to the problem. We hope that the path
integral approach will be more adequate in dealing with the problem of
computing the reduced density matrix or to obtain the master equation for
the atom as is the case in Caldeira-Legget type models \cite%
{feynman,caldeira,paz}. As in previous works, because its exact
integrability, the dressed coordinates were introduced in the context of a
harmonic oscillator coupled linearly to a massless scalar field.
%%%%%%%%%%%%%%%%%%%%%%%%%%%%%%%%%%%%%%%%%%%%%%%%%%%%%%%%%%%%%%%%%%%%%%%%%%%%%%%%%%%%%

The paper is organized as follows: In section 2 we introduce the model and
compute the propagator by exact diagonalization. Section 3 is devoted to
establish the dressed coordinates. In section 4 we compute the probabilities
associated to some physical processes and extend the sum rules found in \cite%
{sumrules}. Finally, in section 5 we give our concluding remarks. Through
this paper we use natural units $c=\hbar =1$.
%%%%%%%%%%%%%%%%%%%%%%%%%%%%%%%%%%%%%%%%%%%%%%%%%%%%%%%%%%%%%%%%%%%%%%%%%%%%%%%%%%%%%
%%%%%%%%%%%%%%%%%%%%%%%%%%%%%%%%%%%%%%%%%%%%%%%%%%%%%%%%%%%%%%%%%%%%%%%%%%%%%%%%%%%%%

\section{The model and its exact diagonalization}

We consider as a toy model of an atom-electromagnetic field system the
system composed by a harmonic oscillator (the atom) coupled to a massless
scalar field. By considering the dipole approximation and expanding in the
field modes we get the following Hamiltonian \cite{adolfo1}
\begin{equation}
H=\frac{1}{2}\left( p_{0}^{2}+\omega _{0}^{2}q_{0}^{2}\right) +\frac{1}{2}%
\sum_{k=1}^{N}\left( p_{k}^{2}+\omega _{k}^{2}q_{k}^{2}\right)
-q_{0}\sum_{k=1}^{N}c_{k}q_{k}+\frac{1}{2}\sum_{k=1}^{N}\frac{c_{k}^{2}}{%
\omega _{k}^{2}}q_{0}^{2}\;,  \label{eq-1}
\end{equation}%
where $q_{0}$ is the oscillator coordinate and $q_{k}$ are the field modes
with $k=1,2,...$; $\omega _{k}=2\pi /L$, $c_{k}=\eta \omega _{k}$, $\eta =%
\sqrt{2g\Delta \omega }$, $\Delta \omega =\omega _{k+1}-\omega _{k}=2\pi /L$%
. With $g$ being a frequency dimensional coupling constant and $L$
the diameter of the sphere in which we confine the oscillator-field
system. In Eq. (\ref{eq-1}) the limit $N\rightarrow \infty $ is to
be understood. The last term in Eq. (\ref{eq-1}) can be seen as a
frequency renormalization \cite{adolfo1} and, it guarantees a
positive-defined Hamiltonian. Due to the Hamiltonian (\ref{eq-1}) is
quadratic in the momenta and there are not constraints we can write
the propagator for the system as being
\begin{equation}
K(\vec{q}_{f},t;\vec{q}_{i},0)=\int \mathcal{D}q_{0}\,\prod_{k=1}^{N}%
\mathcal{D}q_{k}\exp \left( i\int_{0}^{t}dt~L\right) \;,  \label{eq-1a}
\end{equation}%
where ${{\vec{q}}}=\left( q_{0},q_{1},\ldots ,q_{N-1},q_{N}\right) ^{T}$ and
$L$, the Lagrangian, is given by
\begin{eqnarray}
L &=&\frac{1}{2}\left( \dot{q}_{0}^{2}-\bar{\omega}_{0}^{2}q_{0}^{2}\right) +%
\frac{1}{2}\sum_{k=1}^{N}\left( \dot{q}_{k}^{2}-\omega
_{k}^{2}q_{k}^{2}+2q_{0}c_{k}q_{k}\right)  \notag \\
&=&\frac{1}{2}{\dot{\vec{q}}}^{T}{\dot{\vec{q}}}-\frac{1}{2}{\vec{q}}^{T}{%
\mathbf{A}}{\vec{q}}\;,  \label{lag-q}
\end{eqnarray}%
where
\begin{equation}
\bar{\omega}_{0}^{2}=\omega _{0}^{2}+\sum_{k=1}^{N}\frac{c_{k}^{2}}{\omega
_{k}^{2}}\quad ,\qquad  \label{eq-ad}
\end{equation}%
and $\mathbf{A}$ is a symmetric matrix whose components are given by
\begin{equation}
\mathbf{A}=\left(
\begin{array}{ccccccc}
\bar{\omega}_{0}^{2} & -c_{1} & -c_{2} &  &  & -c_{N-1} & -c_{N} \\
-c_{1} & \omega _{1}^{2} &  &  &  &  &  \\
-c_{2} &  & \omega _{2}^{2} &  &  &  &  \\
\vdots &  &  & \ddots &  &  &  \\
\vdots &  &  &  & \ddots &  &  \\
-c_{N-1} &  &  &  &  & \omega _{N-1}^{2} &  \\
-c_{N} &  &  &  &  &  & \omega _{N}^{2}%
\end{array}%
\right) \;.
\end{equation}%
To compute the propagator, Eq. (\ref{eq-1a}), we introduce the coordinate
transformation
\begin{equation}
\vec{q}=\mathbf{T}{\vec{Q}}\qquad ,\qquad {q}_{\mu }=\sum_{r=0}^{N}~t_{\mu
}^{r}Q_{r}\qquad ,\qquad {Q}_{r}=\sum_{\mu =0}^{N}~t_{\mu }^{r}{q}_{\mu }
\label{eq-10}
\end{equation}%
where $\mathbf{T}$\footnote{The components of matrix $\mathbf{T}$
and some of its properties are given in the Appendix. As \textbf{T}
diagonalizes the matrix \textbf{A}, the coordinates $Q_{r}$ are
known as \emph{normal coordinates}} is an orthogonal matrix that
diagonalize $\mathbf{A}$,
\begin{equation}
\mathbf{D}=\mathbf{T}^{T}\!\mathbf{AT}=\mbox{diag}\left( \Omega
_{0}^{2},\Omega _{1}^{2},\Omega _{2}^{2},\cdots ,\Omega _{N-1}^{2},\Omega
_{N}^{2}\right) \;.  \label{eq-10a}
\end{equation}%
It is easy to show \cite{rudnei} that the eigenvalues of $\mathbf{A}$, $%
\Omega _{r}$ are obtained by solving the equation
\begin{equation}
\omega _{0}^{2}-{\Omega }^{2}=\eta ^{2}\sum_{k=1}^{N}\frac{\Omega ^{2}}{%
\omega _{k}^{2}-\Omega ^{2}}\;.  \label{eq10-b}
\end{equation}%
It has shown that such equation has definite positive frequencies
$\Omega ^{2}$ as solutions \cite{adolfo1}. The solutions of the
characteristic equation (\ref{eq10-b}) were used to describe
radiation process in small cavities \cite{adolfo2,gabriel} in good
agreement with the experiment. In \cite{gabriel} this system is also
used to describe a Brownian particle coupled to an ohmic
environment.

Replacing Eq. (\ref{eq-10}) in Eq. (\ref{eq-1a}) we get
\begin{equation}
K(\vec{q}_{f},t;\vec{q}_{i},0)=\prod_{r=0}^{N}\int \mathcal{D}Q_{r}\exp %
\left[ i\int_{0}^{t}\!\!dt\;\left( \frac{1}{2}\dot{Q}_{r}^{2}-\frac{1}{2}%
\Omega _{r}^{2}Q_{r}^{2}\right) \right] \;.  \label{eq-10c}
\end{equation}%
Note that in Eq. (\ref{eq-10c}) the functional measure is maintained, this
because $\det \mathbf{T}=1$. By using the known result for the propagator of
a harmonic oscillator we get for Eq. (\ref{eq-10c}),
\begin{equation}
K(\vec{q}_{f},t;\vec{q}_{i},0)=\prod_{\mu =0}^{N}\left( \frac{\Omega _{\mu }%
}{i2\pi \sin \left( \Omega _{\mu }t\right) }\right) ^{\frac{1}{2}}\,\exp %
\left[ \frac{i\,\Omega _{\mu }}{2\sin \left( \Omega _{\mu }t\right) }\left( %
\left[ Q_{f\,\mu }^{2}+Q_{i\,\mu }^{2}\right] \cos \left( \Omega _{\mu
}t\right) -2Q_{i\,\mu }Q_{f\,\mu }\right) \right] \;.  \label{eq-14x}
\end{equation}

The spectral function is defined as being
\begin{equation}
Y(t)=\int dq_{0}dq_{1}\cdots dq_{N}\,K(\vec{q},t;\vec{q},0)
\end{equation}%
which is easily computed expressing the integral in normal coordinates $%
\left\{ Q_{r}\right\} $, thus we obtain
\begin{equation}
Y(t)=\prod_{r=0}^{N}\left( i2\sin \left( \frac{\Omega _{r}t}{2}\right)
\right) ^{-1}=\sum_{n_{0},...,n_{N}=0}^{\infty }e^{-itE_{n_{0},...,n_{N}}}
\label{eq-14}
\end{equation}%
with the energy spectrum being given by
\begin{equation}
E_{n_{0},...,n_{N}}=\sum_{r=0}^{N}\Omega _{r}\left( n_{r}+\frac{1}{2}\right)
\;.  \label{eq-15}
\end{equation}%
In the $\left\{ Q_{r}\right\} $ coordinates, the {ground state} wave
function is computed from the propagator given by Eq. (\ref{eq-14x}) by
taking the limit $t\rightarrow -i\infty $
\begin{equation}
K(\vec{Q}_{f},t\rightarrow -i\infty ;\vec{Q}_{i},0)=\prod_{r=0}^{N}\left(
\frac{\Omega _{r}}{\pi }\right) ^{\frac{1}{2}}\,\exp \left( -\frac{\,\Omega
_{r}}{2}\left[ Q_{i\,r}^{2}+Q_{f\,r}^{2}\right] -it\frac{\,\Omega _{r}}{2}%
\right) \;,
\end{equation}%
thus the wave function of the ground state is
\begin{equation}
\psi _{00...0}\left( \vec{Q}\right) =\prod_{r=0}^{N}\left( \frac{\Omega _{r}%
}{\pi }\right) ^{\frac{1}{4}}\exp \left( -\frac{\,1}{2}\Omega
_{r}Q_{r}^{2}\right) \;.  \label{e13}
\end{equation}%
We can write the ground state eigenfunction in the original coordinates $%
q_{\mu }$, by using the third equation of (\ref{eq-10})
\begin{equation}
\psi _{00...0}(\vec{q})=\left( \frac{\Omega _{0}}{\pi }\right) ^{\frac{1}{4}%
}\left( \frac{\Omega _{1}}{\pi }\right) ^{\frac{1}{4}}\cdots \left( \frac{%
\Omega _{N}}{\pi }\right) ^{\frac{1}{4}}\exp \left( -\frac{\,1}{2}\sum_{\mu
,\nu =0}^{N}\sum_{r=0}^{N}\Omega _{r}t_{\mu }^{r}t_{\nu }^{r}\ q_{\mu
}q_{\nu }\right) \;.
\end{equation}

\section{The dressed coordinates}

We have observed the vacuum wave function in the $\{q_{\mu }\}$ and normal $%
\left\{ Q_{r}\right\} $ coordinates and we make a question: Is it
possible to find some new set of coordinates $\{\bar{q}_{\mu }\}$ \
relate to them what allow us to describe the oscillators with their
non interacting characteristics? The answer is yes. In such one new
set of coordinates the vacuum wave function must be given as
\begin{equation}
\psi _{00..0}(\bar{q}_{0},...,\bar{q}_{N})=cte.\,\exp \left( -\frac{1}{2}%
\sum_{\mu =0}^{N}\,\omega _{\mu }\left( \bar{q}_{\mu }\right) ^{2}\right)
\end{equation}%
and we named the set of coordinates $\left\{ \bar{q}_{\mu }\right\}
$ as \emph{dressed coordinates}\footnote{The dressed coordinates are
a new type of collective coordinates, which are defined from the
normal coordinates, that allows a correct description for the
absorption and radiation phenomena for a given physical system.}
which describe the oscillators as being non interacting. In a
physical situation we can imagine the atom interacting with a bath
of field modes, i.e., an electromagnetic field. The experience tell
us that the atom does not change his energy spectrum (the energy
spectrum when it is isolated) and only transitions between the
energy levels are observed (absorption and emission process).

Thus we would have for the vacuum state wave functions the following
relation
\begin{equation}
\exp\left(-\frac{1}{2}\sum_{r=0}^{N}\,\Omega_{r}\left(Q_{r}\right)^{2}
\right) \propto \exp \left( -\frac{1}{2}\sum_{\mu =0}^{N}\omega_{\mu
}\,\left( \bar{q}_{\mu }\right)^{2}\right)  \label{eq-24}
\end{equation}

First we will look for the matrix transformation between the normal $\left\{
Q_{\mu }\right\} $ and dressed coordinates $\left\{ {\bar{q}}_{\mu }\right\}
$, thus we set
\begin{equation}
\vec{{\bar{q}}}=\mathbf{M}\vec{Q}\qquad ,\quad \bar{q}_{\mu
}=\sum_{r=0}^{N}M_{\mu }^{r}Q_{r}  \label{eq-25}
\end{equation}%
the quadratic form (\ref{eq-24}) must be invariant under the linear
transformation (\ref{eq-25}), then, to preserve the quadric form we set
\begin{equation}
\sum_{\mu =0}^{N}\omega _{\mu }M_{\mu }^{\,r}M_{\mu }^{s}=\Omega _{r}\delta
_{rs}  \label{eq-27}
\end{equation}%
and to achieve the matrix $\mathbf{M}$ we use the orthonormal matrix $%
\mathbf{T}$ such that
\begin{equation}
M_{\mu }^{\,r}=\sqrt{\frac{\Omega _{r}}{\omega _{\mu }}}\;t_{\mu }^{r}
\end{equation}%
we can show that it satisfies the equation (\ref{eq-27}) by using the
orthonormality condition of the matrix $\mathbf{T}$. The determinant of the
matrix $\mathbf{M}$ is shown to be 1,
\begin{equation*}
\det (\mathbf{M})=\sqrt{\frac{\Omega _{0}\Omega _{1}\cdots \Omega _{N}}{%
\omega _{0}\omega _{1}\cdots \omega _{N}}}\det (\mathbf{T})=1
\end{equation*}%
thus the transformation (\ref{eq-25}) preserve the path-integral measure
defining the propagator (\ref{eq-1a}) of the system. And the inverse
transformation $\left\{ Q_{s}\rightarrow \bar{q}_{\mu }\right\} $ is easily
shown to be
\begin{equation}
Q_{s}=\sum_{\mu =0}^{N}\sqrt{\frac{\omega _{\mu }}{\Omega _{s}}}\;t_{\mu
}^{s}\bar{q}_{\mu }  \label{eq-pt-22}
\end{equation}

\bigskip Then, we come back to the functional integral defining the
propagator (\ref{eq-10c}) in terms of the normal coordinates and using the
transformation (\ref{eq-pt-22}) we can write
\begin{equation}
K(\vec{q}_{f},t;\vec{q}_{i},0)=\int \left( \prod_{\mu =0}^{N}\mathcal{D}\bar{%
q}_{\mu }\right) ~\exp \left\{ i\int_{0}^{t}\!\!dt\;\frac{1}{2}\sum_{\mu
,\nu =0}^{N}\left[ \frac{{}}{{}}Z_{\mu \nu }\dot{\bar{q}}_{\mu }\dot{\bar{q}}%
_{\nu }-C_{\mu \nu }\omega _{\mu }\omega _{\nu }\bar{q}_{\mu }\bar{q}_{\nu }%
\right] \right\}
\end{equation}%
from which we can see the Lagrangian in terms of dressed coordinates is
given by%
\begin{equation}
L_{d}=\frac{1}{2}\sum_{\mu ,\nu =0}^{N}\left[ \frac{{}}{{}}Z_{\mu \nu }\dot{%
\bar{q}}_{\mu }\dot{\bar{q}}_{\nu }-C_{\mu \nu }\omega _{\mu }\omega _{\nu }%
\bar{q}_{\mu }\bar{q}_{\nu }\right]
\end{equation}%
where we have defined the matrices $Z_{\mu \nu }$ and $C_{\mu \nu }$
\begin{equation}
Z_{\mu \nu }=\sqrt{\omega _{\mu }\omega _{\nu }}\sum_{s=0}^{N}\frac{t_{\mu
}^{s}t_{\nu }^{s}}{\Omega _{s}}~\ \ \ ,~\ \ \ \ \ C_{\mu \nu }=\frac{1}{%
\sqrt{\omega _{\mu }\omega _{\nu }}}\sum_{s=0}^{N}\Omega _{s}~t_{\mu
}^{s}t_{\nu }^{s}  \label{cmn}
\end{equation}%
such that
\begin{equation}
Z_{\mu \beta }C_{\beta \nu }=\delta _{\mu \nu }
\end{equation}%
They play the role of renormalization constants and the coordinates $\left\{
\bar{q}_{\mu }\right\} $ are the renormalized coordinates such as it
happened in a renormalized field theory.

To construct the dressed Hamiltonian we first define the dressed momentum $%
\bar{p}_{\mu }$ canonically conjugate to the dressed coordinate $\bar{q}%
_{\mu }$, thus%
\begin{equation}
\bar{p}_{\mu }=\frac{\partial L_{d}}{\partial \dot{\bar{q}}_{\mu }}%
=\sum_{\nu =0}^{N}Z_{\mu \nu }\dot{\bar{q}}_{\nu }
\end{equation}%
from which we get%
\begin{equation}
\dot{\bar{q}}_{\mu }=\sum_{\nu =0}^{N}C_{\mu \nu }\bar{p}_{\nu }
\end{equation}

Thus, the dressed Hamiltonian $H_{d}$ is computed to be%
\begin{equation}
H_{d}=\sum_{\mu ,\nu =0}^{N}\frac{1}{2}C_{\mu \nu }\left( \frac{{}}{{}}\bar{p%
}_{\mu }\bar{p}_{\nu }+\omega _{\mu }\omega _{\nu }\bar{q}_{\mu }\bar{q}%
_{\nu }\right)
\end{equation}

From (\ref{eq-14x}) we write the propagator in dressed coordinates
\begin{eqnarray}
K\left( \bar{q}_{f},T;\bar{q}_{i},0\right) &=&\prod_{r=0}^{N}\left( \frac{%
\omega _{r}}{i2\pi \sin \left( \Omega _{r}T\right) }\right) ^{\frac{1}{2}}
\label{prop-dressed} \\
&&\qquad \times \exp \left( \frac{i}{2}\sum_{s=0}^{N}\sum_{\mu ,\nu =0}^{N}%
\frac{\sqrt{\omega _{\mu }\omega _{\nu }}\,t_{\mu }^{s}t_{\nu }^{s}}{\sin
\left( \Omega _{s}T\right) }\left[ \frac{{}}{{}}\left( \bar{q}_{f\mu }\bar{q}%
_{f\nu }+\bar{q}_{i\mu }\bar{q}_{i\nu }\right) \cos \left( \Omega
_{s}T\right) -2\bar{q}_{f\mu }\bar{q}_{i\nu }\right] \right)  \notag
\end{eqnarray}

It is simple to show that the spectral function computed in dressed
coordinates is the same computed in the equation (\ref{eq-14}), thus, the
energy spectrum (\ref{eq-15}) remains invariant, as it was expected.

The ground state reads as
\begin{equation}
\psi _{00..0}\left( \bar{q}\right) =\left( \frac{\omega _{0}}{\pi }\right)
^{1/4}...\left( \frac{\omega _{N}}{\pi }\right) ^{1/4}\exp \left( -\frac{1}{2%
}\sum_{\alpha =0}^{N}\omega _{\alpha }\left( \bar{q}_{\alpha }\right)
^{2}\right)
\end{equation}
The Hamiltonian operator in dressed coordinates is expressed as
\begin{equation}
H\left( \bar{q}\right) =\sum_{\mu ,\nu =0}^{N}C_{\mu\nu}\left( -\frac{1}{2}%
\frac{\partial }{\partial \bar{q}_{\nu }}\frac{\partial }{\partial \bar{q}%
_{\mu }}+\frac{1}{2}\omega _{\mu }\omega _{\nu }\bar{q}_{\mu }\bar{q}_{\nu
}\right).
\end{equation}
with the coefficients $C_{\mu\nu}$ are given in (\ref{cmn}).

\section{Computing the transition probabilities}

In this section we show what to compute the transition amplitudes of the
system using the exact dressed propagator (\ref{prop-dressed}), thus, we are
interested in the following quantities,
\begin{equation}
\mathcal{A}_{m_0m_1...m_N}^{n_0n_1...n_N}(t)=~_d \langle
n_0,n_1,...,n_N|e^{-iHt}|m_0,m_1,...,m_N\rangle_d\;,  \label{b1}
\end{equation}
that represents the probability amplitude of the system, initially prepared
in the state $|m_0,m_1,...,m_N\rangle$, to be found at time $t$ in the state
$|n_0,n_1,...,n_N\rangle$.

Eq. (\ref{b1}) can be written in terms of the propagator as
\begin{equation}
\mathcal{A}_{m_0m_1...m_N}^{n_0n_1...n_N}(t)=\int d\chi d\xi ~_d\langle
n_0,n_1,...,n_N|\chi\rangle K(\chi,t;\xi,0)\langle \xi
|m_0,m_1,...,m_N\rangle_d\;,  \label{b2}
\end{equation}
where
\begin{equation}
\langle
\xi|m_0,m_1,...,m_N\rangle_d=\psi_{m_0m_1...m_N}(\xi^{\prime}(\xi))\;.
\label{b3}
\end{equation}
Using Eq. (\ref{b3}) we can write Eq. (\ref{b2}) as
\begin{equation}
\mathcal{A}_{m_0m_1...m_N}^{n_0n_1...n_N}(t)= \int d\chi d\xi~\!
\psi_{n_0n_1...n_N}(\chi^{\prime}(\chi)) K(\chi,t;\xi,0)
\psi_{m_0m_1...m_N}(\xi^{\prime}(\xi))  \label{b4}
\end{equation}

First we compute $\mathcal{A}_{0...0m_{\mu }0...0}^{0...0n_{\nu }0...0}(t)$.
Substituting $\psi _{0...0n_{\nu }0...0}(\chi ^{\prime }(\chi ))$, $K(\chi
,t;\xi ,0)$ and $\psi _{0...0m_{\mu }0...0}(\xi ^{\prime }(\xi ))$ we have
\begin{equation}
\mathcal{A}_{0...0m_{\mu }0...0}^{0...0n_{\nu }0...0}(t)=\left[
\prod_{r=0}^{N}\frac{1}{\pi \sqrt{2i\sin (\Omega _{r}t)}}\right] \int d\chi
\frac{H_{n_{\nu }}\left( \sum_{r=0}^{N}t_{\nu }^{r}\chi _{r}\right) }{\sqrt{%
2^{n_{\nu }}n_{\nu }!}}\exp \left( {\frac{i}{2}\sum_{r=0}^{N}\frac{%
e^{i\Omega _{r}t}}{\sin (\Omega _{r}t)}\chi _{r}^{2}}\right) F_{\mu }(\chi
,t)  \label{b5}
\end{equation}%
where
\begin{eqnarray}
F_{\mu }(\chi ,t) &=&\int d\xi \frac{H_{m_{\mu }}\left( \sum_{s=0}^{N}t_{\mu
}^{s}\xi _{s}\right) }{\sqrt{2^{m_{\mu }}m_{\mu }!}}\exp \left[ \frac{i}{2}%
\sum_{s=0}^{N}\left( \frac{e^{i\Omega _{s}t}}{\sin (\Omega _{s}t)}\xi
_{s}^{2}-\frac{2\chi _{s}}{\sin (\Omega _{s}t)}\xi _{s}\right) \right]
\notag \\
&=&\frac{H_{m_{\mu }}\left( \sum_{s=0}^{N}t_{\mu }^{s}i\sin (\Omega _{s}t)%
\frac{\partial }{\partial \chi _{s}}\right) }{\sqrt{2^{m_{\mu }}m_{\mu }!}}%
\int d\xi \exp \left[ \frac{i}{2}\sum_{s=0}^{N}\left( \frac{e^{i\Omega _{s}t}%
}{\sin (\Omega _{s}t)}\xi _{s}^{2}-\frac{2\chi _{s}}{\sin (\Omega _{s}t)}\xi
_{s}\right) \right] \;.  \label{b6}
\end{eqnarray}%
Performing the Gaussian integrals in Eq. (\ref{b6}) we obtain
\begin{equation}
F_{\mu }(\chi ,t)=\left[ \prod_{s=0}^{N}\frac{\sqrt{2\pi i\sin (\Omega _{s}t)%
}}{e^{\frac{i}{2}\Omega _{s}t}}\right] \frac{H_{m_{\mu }}\left(
\sum_{s=0}^{N}t_{\mu }^{s}i\sin (\Omega _{s}t)\frac{\partial }{\partial \chi
_{s}}\right) }{\sqrt{2^{m_{\mu }}m_{\mu }!}}\exp \left( -{\frac{i}{2}%
\sum_{r=0}^{N}\frac{e^{-i\Omega _{r}t}}{\sin (\Omega _{r}t)}\chi _{r}^{2}}%
\right) \;.  \label{b7}
\end{equation}%
Using the identity
\begin{equation}
H_{n}(\sum_{r=0}^{N}t_{\mu }^{r}X_{r})=n!\sum_{+l=n}\frac{(t_{\mu
}^{0})^{l_{0}}}{l_{0}!}\frac{(t_{\mu }^{1})^{l_{1}}}{l_{1}!}...\frac{(t_{\mu
}^{N})^{l_{N}}}{l_{N}!}H_{l_{0}}(X_{0})H_{l_{1}}(X_{1})...H_{l_{N}}(X_{N})
\label{b8}
\end{equation}%
(that holds because $\{t_{\mu }^{r}\}$ is an orthogonal matrix) where $%
+l=(l_{0}+l_{1}+...+l_{N})$ and replacing Eq. (\ref{b7}) in Eq. (\ref{b5})
we get
\begin{equation}
\mathcal{A}_{0...0m_{\mu }0...0}^{0...0n_{\nu }0...0}(t)=\pi ^{-(N+1)/2}e^{-%
\frac{i}{2}\sum_{r=0}^{N}\Omega _{r}t}\sqrt{\frac{m_{\mu }!n_{\nu }!}{%
2^{m_{\mu }+n_{\nu }}}}\sum_{+l=n_{\nu }}\sum_{+s=m_{\mu }}\frac{(t_{\mu
}^{0})^{l_{0}+s_{0}}}{l_{0}!s_{0}!}\frac{(t_{\mu }^{1})^{l_{1}+s_{1}}}{%
l_{1}!s_{1}!}...\frac{(t_{\mu }^{N})^{l_{N}+s_{N}}}{l_{N}!s_{N}!}%
I_{l_{0}s_{0}}I_{l_{1}s_{1}}...I_{l_{N}s_{N}}  \label{b9}
\end{equation}%
where
\begin{eqnarray}
I_{l_{r}s_{r}} &=&\int d\chi _{r}\exp \left( \frac{i}{2}\frac{e^{i\Omega
_{r}t}}{\sin (\Omega _{r}t)}\chi _{r}^{2}\right) H_{l_{r}}(\chi
_{r})H_{s_{r}}\left( i\sin (\Omega _{r}t)\frac{\partial }{\partial \chi _{r}}%
\right) \exp \left( -\frac{i}{2}\frac{e^{-i\Omega _{r}t}}{\sin (\Omega _{r}t)%
}\chi _{r}^{2}\right)  \notag \\
&=&\int d\chi _{r}e^{-\chi _{r}^{2}}H_{l_{r}}(\chi _{r})\left[ \exp \left(
\frac{i}{2}\frac{e^{-i\Omega _{r}t}}{\sin (\Omega _{r}t)}\chi
_{r}^{2}\right) H_{s_{r}}\left( i\sin (\Omega _{r}t)\frac{\partial }{%
\partial \chi _{r}}\right) \exp \left( -\frac{i}{2}\frac{e^{-i\Omega _{r}t}}{%
\sin (\Omega _{r}t)}\chi _{r}^{2}\right) \right] \;.  \label{b10}
\end{eqnarray}

If instead of integrating over coordinates $\xi$ in Eq. (\ref{b5}) we first
integrate over coordinates $\chi$ we would get an expression similar to the
one given in Eq. (\ref{b9}) but with $I_{l_rs_r}$ replaced with $%
I^{\prime}_{l_rs_r}$:
\begin{equation}
I^{\prime}_{l_rs_r}=\int d\xi_r e^{-\chi_r^2}H_{s_r}(\xi_r) \left[\exp\left(%
\frac{i}{2}\frac{e^{-i\Omega_r t}}{\sin(\Omega_r t)}\xi_r^2\right)
H_{l_r}\left(i\sin(\Omega_r t)\frac{\partial}{\partial \xi_r}\right)
\exp\left(-\frac{i}{2}\frac{e^{-i\Omega_r t}}{\sin(\Omega_r t)}\xi_r^2\right)%
\right]\;.  \label{b11}
\end{equation}
Then, since the final result must not depend of the order in which we
perform the integrations we must have $I_{l_rs_r}=I^{\prime}_{l_rs_r}$, and
from Eqs. (\ref{b10}) and (\ref{b11}) we conclude that $%
I_{l_rs_r}=I_{s_rl_r} $.

To perform the integral given in Eq. (\ref{b10}) we have to use the
following theorem
\begin{equation}
\mathrm{if}~k<n\Longrightarrow~\int dx e^{-x^2}H_n(x)x^k=0\;.  \label{b12}
\end{equation}
Note that the expression in brackets in Eq. (\ref{b10}) is a polynomial of
degree $s_r$ in $\xi_r$. Now, if $l_r>s_r$, then by using theorem (\ref{b12}%
), we get $I_{l_rs_r}=0$. Because $I_{l_rs_r}=I_{s_rl_r}$ we also get a
vanishing result for $l_r<s_r$. Then, the only non vanishing result is
obtained for $l_r=s_r$. Using again theorem (\ref{b12}) we note that the
only non vanishing term of the polynomial in brackets is the one of highest
power. Since the highest power of $H_n(x)$ is given by $2^nx^n$ we have for
Eq. (\ref{b10})
\begin{eqnarray}
I_{l_rs_r}&=&[2i\sin(\Omega_r t)]^{s_r} \int d\chi_r
e^{-\chi_r^2}H_{l_r}(\chi_r) \left[\exp\left(\frac{i}{2}\frac{e^{-i\Omega_r
t}}{\sin(\Omega_r t)}\chi_r^2\right) \frac{\partial^{s_r}}{\partial
\chi_r^{s_r}} \exp\left(-\frac{i}{2}\frac{e^{-i\Omega_r t}}{\sin(\Omega_r t)}%
\chi_r^2\right)\right]  \notag \\
&=&e^{-is_r\Omega_r t}\int d\chi_r
e^{-\chi_r^2}H_{l_r}(\chi_r)(2)^{s_r}\chi_r^{s_r}  \notag \\
&=&e^{-is_r\Omega_r t}\int d\chi_r
e^{-\chi_r^2}H_{l_r}(\chi_r)H_{s_r}(\chi_r)  \notag \\
&=&\sqrt{\pi}e^{-is_r\Omega_r t}2^{s_r}s_r!\delta_{l_rs_r}\;.  \label{b13}
\end{eqnarray}
Using Eq. (\ref{b13}) in Eq. (\ref{b9}) we get
\begin{eqnarray}
\mathcal{A}_{0...0m_\mu0...0}^{0...0n_\nu 0...0}(t)&=& e^{-\frac{i}{2}%
\sum_{r=0}^N\Omega_r t} \sqrt{\frac{m_\mu! n_\nu!}{2^{m_\mu+n_\nu}}}
\sum_{+l=m_\mu=n_\nu} \frac{ ( (t_\mu^0)^2 e^{-i\Omega_0} )^{l_0}}{l_0!}
\frac{((t_\mu^1)^2 e^{-i\Omega_1})^{l_1}}{l_1!}... \frac{((t_\mu^N)^2
e^{-i\Omega_N})^{l_N}}{l_N!}2^{l_0+l_1+...+l_N}  \notag \\
&=&e^{-\frac{i}{2}\sum_{r=0}^N\Omega_r t}\delta_{m n}
\left(\sum_{r=0}^Nt_\mu^r t_\nu^re^{-i\Omega_r t}\right)^{n}\;,  \label{b14}
\end{eqnarray}
where in passing to the last line we have used the identity
\begin{equation}
\left(\sum_{r=0}^N X_r\right)^n=n!\sum_{+l=n}\frac{X_0^{l_0}}{l_0!}\frac{%
X_1^{l_1}}{l_1!} ...\frac{X_N^{l_N}}{l_N!}\;.  \label{b15}
\end{equation}
In terms of
\begin{equation}
f_{\mu\nu}(t)= \sum_{r=0}^Nt_\mu^r t_\nu^re^{-i\Omega_r t}\;,  \label{b16}
\end{equation}
Eq. (\ref{b14}) can be written as
\begin{equation}
\mathcal{A}_{0...0m_\mu0...0}^{0...0n_\nu 0...0}(t)=e^{-\frac{i}{2}%
\sum_{r=0}^N\Omega_r t} \delta_{m n}\left[f_{\mu\nu}(t)\right]^n\;.
\label{b17}
\end{equation}

It is straightforward to establish the following identity:
\begin{equation}
\sum_{\mu=0}^N |f_{\nu\mu}(t)|^2=1  \label{c1}
\end{equation}
The proof of the above identity follows trivially by using the
orthonormality property of the matrix $\{t_\mu^r\}$. Writing Eq. (\ref{c1})
for indexes $0$ and $k$ we have
\begin{equation}
|f_{00}(t)|^2+\sum_{k=1}^N |f_{0k}(t)|^2=1\;,  \label{c2}
\end{equation}
\begin{equation}
|f_{k_10}(t)|^2+\sum_{k_2=1}^N |f_{k_1k_2}(t)|^2=1\;.  \label{c3}
\end{equation}

The physical interpretation for the equations above is given as it follows.
Let the initial state of the system given by $|n,1_{k_{1}},1_{k_{2}}...%
\rangle $, the atom in the $n$-th excited level and field quanta of
frequencies $\omega _{k_{1}}$, $\omega _{k_{2}}$, etc. The probability of
this initial states to be found in a measurement performed at time $t$ in
the state $|m,1_{k_{1}^{\prime }},1_{k_{2}^{\prime }}...\rangle $ is denoted
by $\mathcal{P}_{n;1_{k_{1}}1_{k_{2}}...}^{m;1_{k_{1}^{\prime
}}1_{k_{2}^{\prime }}...}(t)$. We know that $\mathcal{P}%
_{1;0}^{1;0}(t)=|f_{00}(t)|^{2}$ is the probability of the oscillator to
remain in the first excited level and $P_{1;0}^{0;1_{k}}(t)=|f_{0k}(t)|^{2}$
is the probability of the oscillator to decay from the first excited level
to the ground state by emission of a field quanta of frequency $\omega _{k}$%
. Obviously in this case we have
\begin{equation}
\mathcal{P}_{1;0}^{1;0}(t)+\sum_{k}P_{1;0}^{0;1_{k}}(t)=1  \label{c4}
\end{equation}%
that is nothing but Eq. (\ref{c2}). Also Eq. (\ref{c3}) can be written as
\begin{equation}
\mathcal{P}_{0;k_{1}}^{1;0}(t)+\sum_{k_{2}}\mathcal{P}%
_{0;k_{1}}^{0;k_{2}}(t)=1\;,  \label{c5}
\end{equation}%
where
\begin{equation}
\mathcal{P}_{0;k_{1}}^{1;0}(t)=|f_{k_{1}0}(t)|^{2}  \label{c5a}
\end{equation}%
and
\begin{equation}
\mathcal{P}_{0;k_{1}}^{0;k_{2}}(t)=|f_{k_{1}k_{2}}(t)|^{2}\;.  \label{c5b}
\end{equation}%
With these identifications the physical meaning of Eq. (\ref{c2}) is clear:
if initially we have a photon of frequency $\omega _{k_{1}}$ and the
oscillator is in its ground state, then at time $t$, either the oscillator
can go to its first excited level by absorption of the initial photon or can
remain in its ground state scattering the initial photon to other photon of
arbitrary frequency.

Note that in establishing the identities (\ref{c4}) and (\ref{c5}) it is
used only the orthogonality property of the matrix $\{t_\mu^r\}$. Then, it
is natural to ask whether it is possible to compute other probabilities
without doing a direct computation as performed in last section. The answer
is yes. For example, if initially the oscillator is in its second excited
level and there are no photons, at time $t$ it can happening that the
oscillator continues in their second excited level, it can go to their first
excited level by emission of photon of arbitrary frequency $\omega_{k_1}$ or
it can decay to their ground state by emission of two photons of arbitrary
frequencies $\omega_{k_1}$ and $\omega_{k_2}$. The respective probabilities
are denoted by $\mathcal{P}_{2;0}^{2;0}(t)$, $\mathcal{P}%
_{2;0}^{1;1_{k_1}}(t)$ and $\mathcal{P}_{2;0}^{0;1_{k_1}1_{k_2}}(t)$.
Obviously we must have
\begin{equation}
\mathcal{P}_{2;0}^{2;0}(t)+\sum_{k_1}\mathcal{P}_{2;0}^{1;1_{k_1}}(t) +
\sum_{k_1k_2}\mathcal{P}_{2;0}^{0;1_{k_1}1_{k_2}}(t)=1\;.  \label{c6}
\end{equation}
Taking the square of Eq. (\ref{c4}) we find
\begin{equation}
\left(\mathcal{P}_{1;0}^{1;0}(t)\right)^2+2\mathcal{P}_{1;0}^{1;0}(t)
\sum_{k_1}P_{1;0}^{0;1_{k_1}}(t)+
\sum_{k_1k_2}P_{1;0}^{0;1_{k_1}}(t)P_{1;0}^{0;1_{k_2}}(t)=1\;.  \label{c7}
\end{equation}

Identifying Eqs. (\ref{c6}) and (\ref{c7}) we obtain
\begin{eqnarray}
\mathcal{P}_{2;0}^{2;0}(t)&=&\left(\mathcal{P}_{1;0}^{1;0}(t)\right)^2
\notag \\
&=&|f_{00}(t)|^4  \label{c8}
\end{eqnarray}
\begin{eqnarray}
\mathcal{P}_{2;0}^{1;1_{k_1}}(t)&=&2\mathcal{P}%
_{1;0}^{1;0}(t)P_{1;0}^{0;1_{k_1}}(t)  \notag \\
&=&2|f_{00}(t)f_{0k_1}(t)|^2  \label{c9}
\end{eqnarray}
and
\begin{eqnarray}
\mathcal{P}_{2;0}^{0;1_{k_1}1_{k_2}}(t)&=&
P_{1;0}^{0;1_{k_1}}(t)P_{1;0}^{0;1_{k_2}}(t)  \notag \\
&=&|f_{0k_1}(t)f_{0k_2}(t)|^2\;.  \label{c10}
\end{eqnarray}

As a second example we consider the oscillator is in its first excited state
and there is one photon of frequency $\omega_{k_1}$. At time $t$ it can
happen that: the oscillator go to its second excited level by absorbing the
initial photon; or the oscillator remains in its first excited state and the
initial photon is scattered to other photon of arbitrary frequency $%
\omega_{k_2}$; or maybe the oscillator can be decay to its ground state by
emission of a photon of arbitrary frequency $\omega_{k_2}$ and the initial
photon is scattered to other photon of frequency $\omega_{k_3}$. The
respective probabilities are denoted by $\mathcal{P}_{1;1_{k_1}}^{2;0}(t)$, $%
\mathcal{P}_{1;1_{k_1}}^{1;1_{k_2}}(t)$ and $\mathcal{P}%
_{1;1_{k_1}}^{0;1_{k_2}1_{k_3}}(t)$. Then, we must have
\begin{equation}
\mathcal{P}_{1;1_{k_1}}^{2;0}(t)+\sum_{k_2}\mathcal{P}%
_{1;1_{k_1}}^{1;1_{k_2}}(t)+ \sum_{k_2k_3}\mathcal{P}%
_{1;1_{k_1}}^{0;1_{k_2}1_{k_3}}(t)\;.  \label{c11}
\end{equation}
Taking Eq. (\ref{c4}) times Eq. (\ref{c5}) we have

\begin{equation}
\mathcal{P}_{1;0}^{1;0}(t)\mathcal{P}_{0;1_{k_1}}^{1;0}(t)+ \sum_{k_2}\left(%
\mathcal{P}_{1;0}^{1;0}(t)\mathcal{P}_{0;1_{k_1}}^{0;1_{k_2}}(t)+ \mathcal{P}%
_{0;1_{k_1}}^{1;0}(t)\mathcal{P}_{1;0}^{0;1_{k_2}}(t)\right)+ \sum_{k_2k_3}%
\mathcal{P}_{1;0}^{0;1_{k_2}}(t)\mathcal{P}_{0;1_{k_1}}^{0;1_{k_3}}(t) =1\;.
\label{c12}
\end{equation}
From Eqs. (\ref{c11}) and (\ref{c12}) we have

\begin{eqnarray}
\mathcal{P}_{1;1_{k_1}}^{2;0}(t)&=&\mathcal{P}_{1;0}^{1;0}(t)\mathcal{P}%
_{0;1_{k_1}}^{1;0}(t)  \notag \\
&=&|f_{00}(t)f_{0k_1}(t)|^2\;,  \label{c14}
\end{eqnarray}
\begin{eqnarray}
\mathcal{P}_{1;1_{k_1}}^{1;1_{k_2}}(t)&=&\mathcal{P}_{1;0}^{1;0}(t) \mathcal{%
P}_{0;1_{k_1}}^{0;1_{k_2}}(t)+ \mathcal{P}_{0;1_{k_1}}^{1;0}(t)\mathcal{P}%
_{1;0}^{0;1_{k_2}}(t)  \notag \\
&=&|f_{00}(t)f_{k_1k_2}(t)|^2+|f_{0k_1}(t)f_{0k_2}(t)|^2  \label{c15}
\end{eqnarray}
and
\begin{eqnarray}
\mathcal{P}_{1;1_{k_1}}^{0;1_{k_2}1_{k_3}}(t)&=&\frac{1}{2}\left( \mathcal{P}%
_{1;0}^{0;1_{k_2}}(t)\mathcal{P}_{0;1_{k_1}}^{0;1_{k_3}}(t)+ \mathcal{P}%
_{1;0}^{0;1_{k_3}}(t)\mathcal{P}_{0;1_{k_1}}^{0;1_{k_2}}(t)\right)  \notag \\
&=&\frac{1}{2}\left(|f_{0k_2}(t)f_{k_1k_3}|^2+|f_{0k_3}(t)f_{k_1k_2}|^2%
\right)\;.  \label{c16}
\end{eqnarray}

And so we can give all the probabilities associated to any decay or
absorption process placing in the system.

\section{Conclusions}

In the present work is shown using the path-integral formalism that
the dressed coordinates appear as a coordinate transformation
preserving the quadric form that defines the ground state wave
function of the system guaranteeing the vacuum stability and, they
also leave invariant the functional measure of the path-integral.
Within the Hamiltonian formalism it can be shown that such linear
transformation, which defines the dressed coordinates, also leaves
invariant the canonical form of the action. Thus, the dressed
coordinates can be also defined via a canonical transformation.

The calculus of the transition amplitudes has been performed using
the dressed propagator, being obtained the basic formula which
defines the \emph{sum rules} presented in \cite{sumrules}, then, the
rules have been extended to describe other physical processes. In
spite of the computation seems very difficult, the dressed
coordinates allow to use  the properties of the Hermite polynomials
simplifying greatly the calculus.

On the other hand, it has been made an extensive use of the model
given by Eq. (\ref{eq-1}) to study different physical situations,
such as the quantum Brownian motion, decoherence and other related
problems in quantum optics. In such context, it is interesting the
computation of the reduced matrix density for the model (\ref{eq-1})
in the framework of dressed coordinates; the advances in such
direction will be reported elsewhere.

\subsection*{Acknowledgement}

GFH (grant 02/09951-3) and RC (grant 01/12611-7) thank to FAPESP for full
support. BMP thanks CNPq and FAPESP (grant 02/00222-9) for partial support.

\appendix

\section{The orthonormal matrix T=$\left[ t_{\protect\mu }^{s}\right] $}

Because the orthogonal character of the $\mathbf{T}$-matrix its components
satisfy
\begin{equation}
\sum_{\mu =0}^{N}~t_{\mu }^{r}t_{\mu }^{s}=\delta _{rs}~\ \ ,~\ \
\sum_{r=0}^{N}~t_{\mu }^{r}t_{\nu }^{r}=\delta _{\mu \nu }  \label{app-1}
\end{equation}%
and from the Lagrangian (\ref{lag-q}) expressed in terms of the normal $%
\left\{ Q_{r}\right\} $ we get other important relation
\begin{equation}
\sum_{\mu =0}^{N}~\bar{\omega}_{\mu }^{2}\ t_{\mu }^{r}t_{\mu
}^{s}~-\sum_{k=1}^{N}~\eta \omega
_{k}~t_{0}^{r}t_{k}^{s}-\sum_{k=1}^{N}~\eta \omega
_{k}~t_{0}^{s}t_{k}^{r}=\Omega _{r}^{2}\delta _{rs}  \label{app-2}
\end{equation}%
where $\bar{\omega}_{0}^{2}$ have been defined in (\ref{eq-ad}) and $\bar{%
\omega}_{k}^{2}={\omega }_{k}^{2}$ . Using the equations above we can show
the following sum
\begin{equation}
\bar{\omega}_{0}^{2}=\sum_{s=0}^{N}~\Omega _{s}^{2}\left( t_{0}^{s}\right)
^{2}~\ \ \ \ ,~\ \ \ \ \eta \omega _{k}=-\sum_{s=0}^{N}~\Omega
_{s}^{2}t_{k}^{s}t_{0}^{s}~\ \ \ \ ,~\ \ \ \ \omega
_{k}^{2}=\sum_{s=0}^{N}\Omega _{s}^{2}\left( t_{k}^{s}\right) ^{2}
\label{app-3}
\end{equation}%
and also compute the elements of the $\mathbf{T}$-matrix%
\begin{equation}
t_{k}^{s}=\frac{\eta \omega _{k}}{\omega _{k}^{2}-\Omega _{s}^{2}}%
t_{0}^{s}~\ \ \ \ ,~\ \ \ \ t_{0}^{s}=\left[ 1+\eta ^{2}\sum_{k=1}^{N}\frac{%
\omega _{k}^{2}}{\left( \omega _{k}^{2}-\Omega _{s}^{2}\right) ^{2}}\right]
^{-\frac{1}{2}}  \label{app-4}
\end{equation}

\end{document}